\documentclass[conference]{IEEEtran}
\IEEEoverridecommandlockouts
\usepackage{cite}
\usepackage{graphicx}
\usepackage{textcomp}
\usepackage{xcolor}
\usepackage{graphicx}
\usepackage{caption}
\usepackage{subcaption}
\usepackage{hyperref}
\usepackage{amsmath,amssymb,amsfonts,bm}
\usepackage{stmaryrd}
\usepackage{textgreek}
\usepackage{multirow}
\usepackage{algorithm}
\usepackage[noend]{algpseudocode}
\usepackage{ulem}
\usepackage{tikz}
\usepackage{booktabs}
\usepackage{makecell}
\usepackage{varwidth}
\usepackage{enumitem}

\def\BibTeX{{\rm B\kern-.05em{\sc i\kern-.025em b}\kern-.08em
    T\kern-.1667em\lower.7ex\hbox{E}\kern-.125emX}}
\begin{document}

\title{HHEML: Hybrid Homomorphic Encryption for Privacy-Preserving Machine Learning on Edge}

\author{\IEEEauthorblockN{Yu Hin Chan, Hao Yang, Shiyu Shen, Xingyu Fan, Shengzhe Lyu, Patrick S. Y. Hung, Ray C. C. Cheung}}

\maketitle{}

\begin{abstract}
Privacy-preserving machine learning (PPML) is an emerging topic to handle secure machine learning inference over sensitive data in untrusted environments. Fully homomorphic encryption (FHE) enables computation directly on encrypted data on the server side, making it a promising approach for PPML. However, it introduces significant communication and computation overhead on the client side, making it impractical for edge devices. Hybrid homomorphic encryption (HHE) addresses this limitation by combining symmetric encryption (SE) with FHE to reduce the computational cost on the client side, and combining with an FHE-friendly SE can also lessen the processing overhead on the server side, making it a more balanced and efficient alternative. Our work proposes a hardware-accelerated HHE architecture built around a lightweight symmetric cipher optimized for FHE compatibility and implemented as a dedicated hardware accelerator. To the best of our knowledge, this is the first design to integrate an end-to-end HHE framework with hardware acceleration. Beyond this, we also present several microarchitectural optimizations to achieve higher performance and energy efficiency. The proposed work is integrated into a full PPML pipeline, enabling secure inference with significantly lower latency and power consumption than software implementations. Our contributions validate the feasibility of low-power, hardware-accelerated HHE for edge deployment and provide a hardware-software co-design methodology for building scalable, secure machine learning systems in resource-constrained environments. Experiments on a PYNQ-Z2 platform with the MNIST dataset show over a 50× reduction in client-side encryption latency and nearly a 2× gain in hardware throughput compared to existing FPGA-based HHE accelerators.
\end{abstract}


\section{Introduction}
\label{sec:introduction}

Contemporary machine learning (ML) workloads often involve handling sensitive datasets, such as personal health information, biometric authentication data, confidential business analytics, and sensitive financial records~\cite{cryptonets2016,gilad2017lola}. The delivery of ML as a cloud-based or remote inference service necessitates transferring raw data inputs, intermediate representations, or model parameters beyond the direct control of the original data custodians. This inherently poses substantial risks, ranging from accidental exposure to intentional misuse~\cite{rathee2020falcon,wang2021cheetah}. To effectively mitigate these concerns, the research community has increasingly concentrated efforts on privacy-preserving machine learning (PPML)~\cite{podschwadt2022survey,yuan2025approximate}, a field dedicated to maintaining data confidentiality while preserving the full functionality of ML algorithms.

Privacy-preserving machine learning (PPML) encompasses several prominent techniques, notably fully homomorphic encryption (FHE)~\cite{lee2022privacy,cryptonets2016}, differential privacy (DP), and secure multi-party computation based on garbled circuits (GC)~\cite{wang2021cheetah}. Among these methods, FHE is distinguished by its robust security model, enabling arbitrary computations directly on encrypted data without any intermediate decryption~\cite{podschwadt2022survey,yuan2025approximate}. In a typical FHE-based PPML workflow, the client locally encrypts sensitive input data before transmitting the ciphertexts to an untrusted computation server. This server performs inference or training entirely within the encrypted domain, returning encrypted results that can only be decrypted by the client, who exclusively holds the corresponding secret key.

Compared to alternative PPML methods, FHE provides end-to-end cryptographic security without compromising data accuracy or necessitating intensive communication~\cite{lee2022privacy}. In contrast, DP ensures privacy through calibrated statistical noise injection, inherently sacrificing precision~\cite{yuan2025approximate}. Likewise, GC-based methods require frequent interactions among multiple parties, thus limiting practicality in scenarios characterized by unreliable or slow network connections~\cite{wang2021cheetah}. Conversely, FHE-based solutions are inherently non-interactive, involve only a single communication exchange per computation task, and demonstrate robustness under conditions of high latency or intermittent connectivity~\cite{cryptonets2016,gilad2017lola}. These advantages make FHE particularly well-suited to Internet-of-Things (IoT) deployments, where devices often face resource limitations, heterogeneity, and unstable connections while simultaneously requiring rigorous data privacy protection~\cite{frimpong2024guardml}.

Although FHE stands out among existing PPML methods in IoT scenarios, its practical deployment remains challenging. Prominent FHE schemes, such as Brakerski-Gentry-Vaikuntanathan (BGV), Brakerski/Fan-Vercauteren (BFV), Cheon-Kim-Kim-Song (CKKS), Fully Homomorphic Encryption over the Torus (FHEW), and Torus Fully Homomorphic Encryption (TFHE)~\cite{podschwadt2022survey}, typically involve significant computational overhead and extensive memory usage. Each critical phase, including plaintext encoding, encryption, and homomorphic evaluation, demands substantial computational resources, and ciphertexts impose orders-of-magnitude overhead compared to conventional symmetric encryption. While recent advances in algorithms and optimized implementations have progressively improved the performance of FHE schemes~\cite{gong2024practical,samardzic2022craterlake,kim2023sharp,fan2023tensorfhe}, the inherent complexity continues to be prohibitive for ultra-low-power IoT devices that possess limited computational capability, constrained memory, and strict energy budgets~\cite{krieger2024aloha}.

\begin{figure}
    \centering
    \includegraphics[width=1\linewidth]{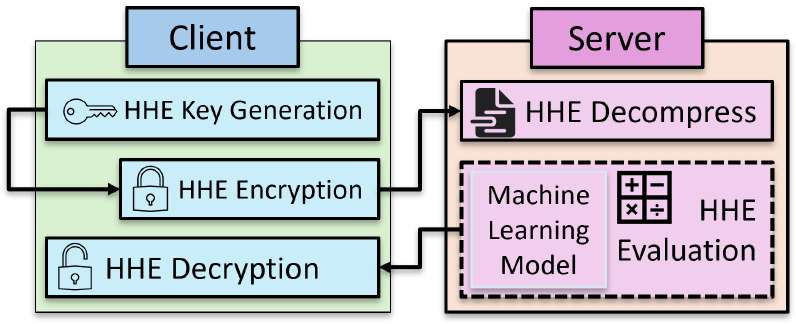}
    \caption{The Workflow of an HHE-based PPML System}
    \label{fig:HHE PPML overview}
\end{figure}

To address these practical challenges, hybrid homomorphic encryption (HHE) has emerged as a promising alternative~\cite{dobraunig2023pasta,masta2024,elisabeth2024,sok2025}. The HHE-based PPML system model is shown in Fig. \ref{fig:HHE PPML overview}. HHE integrates efficient symmetric encryption methods, typically lightweight block ciphers, with the algebraically rich capabilities of FHE. This strategic combination enables end-to-end secure computation while significantly reducing ciphertext expansion and computational overhead~\cite{frimpong2024guardml,nguyen2024pervasive}. Consequently, HHE is particularly advantageous for edge computing scenarios, where performance, bandwidth, and energy constraints must be carefully balanced against rigorous data privacy requirements~\cite{aikata2025pasta,jeon2025presto}.

In light of the limitations associated with classical FHE in edge computing scenarios, we propose HHEML, a hardware accelerated framework for HHE tailored to PPML on resource constrained devices. Built upon the Pasta encryption scheme~\cite{dobraunig2023pasta,aikata2025pasta}, which combines symmetric encryption with homomorphic compatibility, HHEML is designed to reduce latency and energy consumption at the client side. To further alleviate communication overhead between software and hardware components, we implement a streamlined data transfer pipeline that leverages processing system to programmable logic (PS to PL) communication and a lightweight Ethernet-based protocol for exchanging encrypted data with a remote server.

The first contribution of this work is the development of an end-to-end hardware acceleration framework compatible with FHE, implemented on the PYNQ-Z2 FPGA~\cite{aikata2025pasta}. The system includes a dedicated module for symmetric encryption, such as Advanced Encryption Standard (AES)~\cite{gong2024practical}, a high bandwidth PS to PL AXI interface, and a communication layer based on Ethernet for transferring ciphertexts. This configuration allows for efficient data encryption at the edge, followed by secure transformation into FHE ciphertext for remote inference or training.

The second contribution targets the server-side cost of transforming symmetric encryption ciphertexts into FHE format. To address this, we adopt the Pasta cipher~\cite{dobraunig2023pasta}, which preserves algebraic structure during encryption and is optimized for homomorphic applications. We develop a custom hardware module to offload these operations from the server software, thereby improving transformation efficiency and reducing overall processing delay in the PPML pipeline~\cite{frimpong2024guardml}.

The third contribution is the integration of HHEML into a complete PPML inference pipeline and a detailed evaluation of its performance. We validate the system on representative benchmarks and demonstrate substantial improvements in throughput and power efficiency when compared with software-only FHE baselines~\cite{cryptonets2016,gilad2017lola,frimpong2024guardml}. These results underscore the feasibility and impact of hardware accelerated HHE in real world edge computing environments~\cite{jeon2025presto}.

The main contributions of this work are summarized as follows:
\begin{itemize}
    \item The design of an FPGA-based HHE acceleration framework that integrates symmetric encryption and FHE compatibility for edge devices.
    \item A hardware implementation of the Pasta cipher for efficient ciphertext transformation at the server side.
    \item A system-level evaluation within a PPML pipeline, demonstrating improved throughput and energy efficiency over conventional approaches.
\end{itemize}

\section{Related Work}
\label{sec:related_work}

\subsection{Hardware Accelerators of Homomorphic Encryption}

Hardware accelerators are pivotal in enhancing the practicality of homomorphic encryption (HE) by addressing its significant computational overhead. For fully homomorphic encryption (FHE), which supports arbitrary computations on encrypted data, several hardware solutions have been developed. CraterLake~\cite{samardzic2022craterlake} introduces a hardware accelerator that enables unbounded computation on encrypted data by optimizing bootstrapping operations, a critical bottleneck in FHE. REED~\cite{aikata2023reed} presents a chiplet-based architecture for FHE, offering improved scalability and flexibility over monolithic designs. SHARP~\cite{kim2023sharp} proposes a short-word hierarchical accelerator tailored for the CKKS scheme, enhancing robustness and efficiency. Aloha-HE~\cite{krieger2024aloha} provides a low-area hardware accelerator for client-side HE operations, making it suitable for resource-constrained environments. GPU-based accelerators, such as Cheddar~\cite{kim2024cheddar} and Phantom~\cite{yang2024phantom}, leverage parallel processing to accelerate HE computations, while TensorFHE~\cite{fan2023tensorfhe} demonstrates the use of GPGPUs for practical encrypted data processing. A comprehensive survey by~\cite{gong2024practical} analyzes existing acceleration methods for FHE, categorizing them into algorithmic and hardware-based approaches and providing evaluation metrics for comparison.

In the context of hybrid homomorphic encryption (HHE), which combines different HE schemes or integrates HE with other cryptographic techniques to balance security and efficiency, hardware acceleration is equally vital. Pasta~\cite{dobraunig2023pasta} establishes the theoretical foundation for HHE, proposing a symmetric encryption scheme optimized for integer HHE use cases. Building on this, Pasta on Edge~\cite{aikata2025pasta} presents the first implementation of an HHE scheme as a cryptoprocessor on FPGA and ASIC platforms, achieving a 43--171$\times$ speedup on low-end 130nm ASIC technology and a 97$\times$ speedup on high-end 7nm and 28nm ASICs compared to previous FHE accelerators. Similarly, Presto~\cite{jeon2025presto} develops hardware accelerators for HHE ciphers HERA and Rubato on an AMD Virtex UltraScale+ FPGA. Other recent designs such as Masta~\cite{masta2024} and Elisabeth~\cite{elisabeth2024} explore new families of FHE-friendly symmetric ciphers optimized for transciphering operations. The SoK on FHE-friendly symmetric ciphers~\cite{sok2025} provides a comprehensive benchmarking of Pasta, Masta, and Elisabeth under ML workloads such as ResNet-20, outlining their trade-offs in depth, latency, and energy efficiency. These advancements highlight the critical role of hardware solutions in making HHE practical for privacy-preserving computations.

\subsection{Homomorphic Encryption and Deep Learning}

The integration of homomorphic encryption (HE) with deep learning (DL) has gained significant traction for enabling privacy-preserving computations in sensitive domains such as healthcare and finance. Early frameworks like CryptoNets~\cite{cryptonets2016} and LoLa~\cite{gilad2017lola} pioneered encrypted inference using FHE but suffered from high latency and ciphertext expansion. Later works such as Falcon~\cite{rathee2020falcon} and Cheetah~\cite{wang2021cheetah} employed secure multiparty computation or optimized HE techniques to reduce inference time, but communication costs and limited scalability remained bottlenecks. More recent surveys~\cite{podschwadt2022survey,yuan2025approximate} review such FHE-based PPML frameworks, detailing their network adaptations, accuracy trade-offs, and runtime overheads. Lee et al.~\cite{lee2022privacy} propose a privacy-preserving deep neural network inference system using FHE, demonstrating feasibility but highlighting performance constraints. Tong et al.~\cite{tong2024accurate} mitigate these constraints via polynomial approximations for non-linear activations, improving throughput while maintaining accuracy.

Hybrid homomorphic encryption (HHE) has emerged as a promising alternative to reduce client-side costs. GuardML~\cite{frimpong2024guardml} demonstrates practical PPML services by combining lightweight symmetric encryption with FHE transformations, achieving improved inference latency on constrained devices. Nguyen et al.~\cite{nguyen2024pervasive} propose pervasive PPML protocols leveraging HHE for secure computations in zero-trust environments, while Podschwadt et al.~\cite{podschwadt2024memory} address memory limitations of HE-based models for resource-constrained devices. The SoK on FHE-friendly ciphers~\cite{sok2025} further benchmarks hybrid ciphers such as Pasta, Masta, and Elisabeth for secure ML inference, showing significant performance benefits over traditional FHE-only approaches. These developments underscore the promising role of HHE in improving the efficiency and scalability of privacy-preserving ML.

\subsection{Limitations of Previous Work}

Prior work on privacy-preserving machine learning often focuses either on standalone hardware accelerators for HE/HHE without full integration into PPML pipelines or purely software-based solutions that suffer from high client-side latency. Few works consider a complete end-to-end architecture that combines hardware acceleration, hybrid encryption, and efficient communication between client and server. HHEML bridges this gap by introducing a hardware/software co-designed framework with FPGA acceleration, pipelined data processing, and Ethernet-enabled offloading to server-side FHE inference.

\section{Background}
\label{sec:background}

\subsection{Notations}
\label{sec:notations}

Throughout this paper, we use the following notations. Let $\mathbb{F}_p$ denote a finite field of prime order $p$. The internal state of the Pasta cipher is represented as $\mathbf{x} = \mathbf{x}_L \,\Vert\, \mathbf{x}_R$, where each half consists of $t$ words of 32 bits. A symmetric secret key is denoted by $\mathsf{sk}$, while $N$ represents a public nonce and $i$ indexes the input block in a message stream. Plaintext and ciphertext blocks are denoted by $\mathbf{m}_i$ and $\mathbf{c}_i$, respectively. The core permutation of Pasta is $\mathrm{Pasta}\text{-}\pi(\cdot)$, applied over $r$ rounds comprising affine layers $A_{j,N,i}$ and substitution layers $S, S'$. The keystream is expanded using an extendable-output function (XOF, e.g., SHAKE128), and $\mathrm{left}_t(\cdot)$ extracts the first $t$ words from a permutation output. In our hardware design, DMA refers to Direct Memory Access used for high-throughput data transfer between the processing system (PS) and programmable logic (PL).

\subsection{Symmetric and Asymmetric Encryption}
Symmetric and asymmetric encryption form the foundation of modern cryptographic systems~\cite{gong2024practical}, each with distinct operational principles, security properties, and performance characteristics. Symmetric encryption employs a single shared secret key for both encryption and decryption, offering high throughput and low computational cost. It is widely used in environments with constrained resources, such as embedded systems, encrypted storage, and real-time communication~\cite{lee2022privacy}. In contrast, asymmetric encryption is based on a public and private key pair, enabling secure communication without pre-shared keys and supporting functions such as digital signatures and key encapsulation~\cite{gong2024practical}. Although schemes like RSA and elliptic curve cryptography (ECC) provide flexible trust models and strong security guarantees, they are inherently more computationally demanding and less suited for processing large volumes of data. As a result, many cryptographic protocols adopt a hybrid approach in which asymmetric encryption establishes secure channels or performs key exchange, while symmetric encryption secures the data payload~\cite{lee2022privacy}. This design pattern also informs the structure of homomorphic encryption systems, where lightweight symmetric components are increasingly employed to alleviate the computational overhead of public-key operations~\cite{dobraunig2023pasta,frimpong2024guardml}.

\subsection{Homomorphic Encryption}
Homomorphic encryption (HE) enhances traditional cryptographic techniques by enabling computations to be carried out directly on encrypted data~\cite{cryptonets2016,gilad2017lola}. This allows an untrusted party, such as a cloud service provider, to perform meaningful operations without access to the plaintext, thereby preserving data confidentiality throughout the computational process. The fully homomorphic encryption (FHE) model supports arbitrary computations on ciphertexts and represents the most powerful instantiation of HE~\cite{lee2022privacy,podschwadt2022survey}. However, this expressiveness comes at a substantial performance cost. Each stage in the FHE workflow, ranging from encoding and encryption to homomorphic evaluation and decryption, involves complex arithmetic over high-dimensional ciphertexts~\cite{gong2024practical,yuan2025approximate}. As a result, FHE systems typically incur orders-of-magnitude increases in computation time and memory usage, along with significant ciphertext expansion~\cite{samardzic2022craterlake,kim2023sharp}.

Hybrid homomorphic encryption (HHE) has been proposed as a practical alternative to address these challenges~\cite{dobraunig2023pasta,frimpong2024guardml}. By combining efficient symmetric encryption for general data protection with targeted use of FHE for sensitive or security-critical operations, HHE reduces the computational demands placed on the client while preserving strong end-to-end security guarantees~\cite{nguyen2024pervasive,masta2024,elisabeth2024,sok2025}. This balanced design has enabled new applications of privacy-preserving machine learning in edge computing scenarios, where devices operate under strict resource constraints related to performance, bandwidth, and energy consumption~\cite{frimpong2024guardml,aikata2025pasta}.

\subsection{Advanced Encryption Standard (AES)}
The Advanced Encryption Standard (AES), defined in NIST FIPS 197~\cite{gong2024practical}, is a symmetric-key block cipher that operates on 128-bit plaintext blocks and supports key sizes of 128, 192, or 256 bits. The algorithm applies a sequence of transformation rounds to the plaintext, where the number of rounds Nr depends on the key length: 10, 12, or 14 rounds for 128-, 192-, and 256-bit keys, respectively. Each round uses a unique 128-bit round key derived from the original key via the key expansion algorithm~\cite{gong2024practical}.
\begin{algorithm}
\caption{Pseudocode for \textsc {CIPHER()}}\label{alg:cipher}
\begin{algorithmic}[1]
\Procedure{Cipher}{\textit{in}, $N_r$, $w$}
  \State $state \gets in$ 
  \State $state \gets \text{AddRoundKey}(state, w[0..3])$ 
  \For{$round$ from 1 to $N_r - 1$}
    \State $state \gets \text{SubBytes}(state)$ 
    \State $state \gets \text{ShiftRows}(state)$ 
    \State $state \gets \text{MixColumns}(state)$ 
    \State $state \gets \text{AddRoundKey}(state, w[4 \cdot round .. 4 \cdot round + 3])$
  \EndFor
  \State $state \gets \text{SubBytes}(state)$
  \State $state \gets \text{ShiftRows}(state)$
  \State $state \gets \text{AddRoundKey}(state, w[4 \cdot N_r .. 4 \cdot N_r + 3])$ 
  \State \Return $state$
\EndProcedure
\end{algorithmic}
\end{algorithm}

\begin{algorithm}
\caption{Pseudocode for \textsc{InvCipher()}}\label{alg:invcipher}
\begin{algorithmic}[2]
\Procedure{InvCipher}{\textit{in}, $N_r$, $w$}
  \State $state \gets in$ 
  \State $state \gets \text{AddRoundKey}(state, w[4 \cdot N_r .. 4 \cdot N_r + 3])$ 
  \For{$round$ from $N_r - 1$ downto $1$}
    \State $state \gets \text{InvShiftRows}(state)$ 
    \State $state \gets \text{InvSubBytes}(state)$ 
    \State $state \gets \text{AddRoundKey}(state, w[4 \cdot round .. 4 \cdot round + 3])$
    \State $state \gets \text{InvMixColumns}(state)$ 
  \EndFor
  \State $state \gets \text{InvShiftRows}(state)$
  \State $state \gets \text{InvSubBytes}(state)$
  \State $state \gets \text{AddRoundKey}(state, w[0..3])$
  \State \Return $state$
\EndProcedure
\end{algorithmic}
\end{algorithm}
The AES encryption process, defined by the Cipher() procedure, begins with an initial AddRoundKey, followed by $N_r - 1$ rounds consisting of SubBytes, ShiftRows, MixColumns, and AddRoundKey, and ends with a final round excluding MixColumns~\cite{lee2022privacy}. Decryption, via InvCipher(), reverses this sequence: starting with AddRoundKey, applying $N_r - 1$ rounds of InvShiftRows, InvSubBytes, AddRoundKey, and InvMixColumns, and concluding with InvShiftRows, InvSubBytes, and a final AddRoundKey. Both procedures share the same expanded key schedule and ensure cryptographic strength through a combination of substitution, permutation, and key mixing~\cite{gong2024practical}. Each round transforms the internal state as follows:
\begin{itemize}
\item SubBytes: A nonlinear substitution step that replaces each byte in the state using a fixed S-box, providing confusion.
\item ShiftRows: A transposition step where the rows of the state matrix are cyclically shifted by different offsets, contributing to diffusion.
\item MixColumns: A column-wise mixing operation that transforms each column of the state using linear algebra over a finite field, further enhancing diffusion.
\item AddRoundKey: A bitwise XOR between the current state and a round-specific subkey derived from the original cipher key, injecting key material into each round.
\end{itemize}
The inverse operations, including InvSubBytes, InvShiftRows, and InvMixColumns, reverse their corresponding encryption steps~\cite{gong2024practical}.

\subsection{Pasta}
Conventional symmetric ciphers like AES are ill-suited for homomorphic encryption (HE) contexts due to their reliance on nonlinear operations such as S-box lookups, table-driven substitutions, and fine-grained bitwise logic, all of which are expensive to emulate over encrypted data~\cite{lee2022privacy}. In contrast, Pasta is a symmetric cipher specifically designed to align with the computational structure of lattice-based HE schemes~\cite{dobraunig2023pasta,aikata2025pasta,sok2025}. It avoids homomorphically inefficient operations and instead leverages FHE-friendly primitives such as modular addition, modular multiplication, and word-level linear transformations~\cite{dobraunig2023pasta}.

Pasta adopts a substitution-permutation network (SPN)-like architecture operating over 32-bit words~\cite{dobraunig2023pasta}. Its nonlinear components are chosen to minimize multiplicative depth while ensuring adequate cryptographic diffusion, which is critical for FHE efficiency. This design enables practical hybrid homomorphic encryption (HHE) systems, where symmetric encryption is performed on the client side and homomorphic evaluation is offloaded to the server~\cite{frimpong2024guardml}. Its operation can be broken down into three main phases:

\begin{enumerate}[label={\arabic*.}, left=0pt]

  \item {State and Key Schedule}
    \begin{itemize}[topsep=0pt]
      \item The internal state is a vector 
        \[
          \mathbf{x} \in \mathbb{F}_p^{2t}, \quad
          \mathbf{x} = \mathbf{x}_L \,\Vert\, \mathbf{x}_R
        \]
        split into left and right halves.
      \item A secret key 
        \(\mathsf{sk} \in \mathbb{F}_p^{2t}\),
        along with a public nonce \(N\) and block counter \(i\), seeds an extendable-output function (e.g., SHAKE128) to generate round constants and affine-layer matrices.
    \end{itemize}

  \item {Keystream Generation (Encryption/Decryption)}
    \begin{itemize}[topsep=0pt]
      \item For a plaintext block \(\mathbf{m}_i \in \mathbb{F}_p^t\), the ciphertext is computed as
        \[
          \mathbf{c}_i = \mathbf{m}_i + \mathrm{left}_t\left(\mathrm{Pasta}\text{-}\pi(\mathsf{sk}, N, i)\right),
        \]
        where \(\mathrm{left}_t(\cdot)\) extracts the first \(t\) words of the permutation output.
      \item Decryption recovers \(\mathbf{m}_i\) by subtracting the same keystream component from \(\mathbf{c}_i\).
    \end{itemize}

  \item {Core Permutation \(\mathrm{Pasta}\text{-}\pi\)}
    \begin{itemize}[topsep=0pt]
      \item The core permutation consists of \(r\) rounds (e.g., \(r=3\) for Pasta‑3 and \(r=4\) for Pasta‑4), each comprising a linear affine layer and a nonlinear substitution.
      \item The permutation over state \(\mathbf{x} \in \mathbb{F}_p^{2t}\) with nonce \(N\) and block counter \(i\) is defined as:
        \begin{align*}
          \mathrm{Pasta}\text{-}\pi(\mathbf{x}, N, i)
          &= A_{r,N,i} \circ S \circ A_{r-1,N,i} \circ S' \circ \cdots \\
          &\quad \circ A_{1,N,i} \circ S' \circ A_{0,N,i}(\mathbf{x}).
        \end{align*}

      \item Each component is defined as follows:
      \paragraph{Affine Layer \(\boldsymbol{A_{j,N,i}}\).}
      Applies an invertible linear transformation \(M_{j,N,i}\) followed by a round constant, both derived from SHAKE128. This layer ensures reversible diffusion across the state.
    
      \paragraph{Feistel-Style S‑Box \(\boldsymbol{S'}\).}
      Used in rounds \(0\) through \(r{-}1\), this low-depth substitution maps each element by:
      \[
      [S'(\mathbf{x})]_i =
      \begin{cases}
      x_i, & i = 0 \\
      x_i + x_{i-1}^2, & \text{otherwise}
      \end{cases}
      \]
      and is efficiently implementable with a single multiplication and rotation.
    
      \paragraph{Cube S‑Box \(\boldsymbol{S}\).}
      Applied only in the final round, this nonlinearity is defined as
      \[
      S(x_i) = x_i^3,
      \]
      providing stronger algebraic complexity while remaining compatible with FHE operations.    
    \end{itemize}
\end{enumerate}

% algorithm detail

\section{HHEML}

\begin{figure}
    \centering
    \includegraphics[width=1\linewidth]{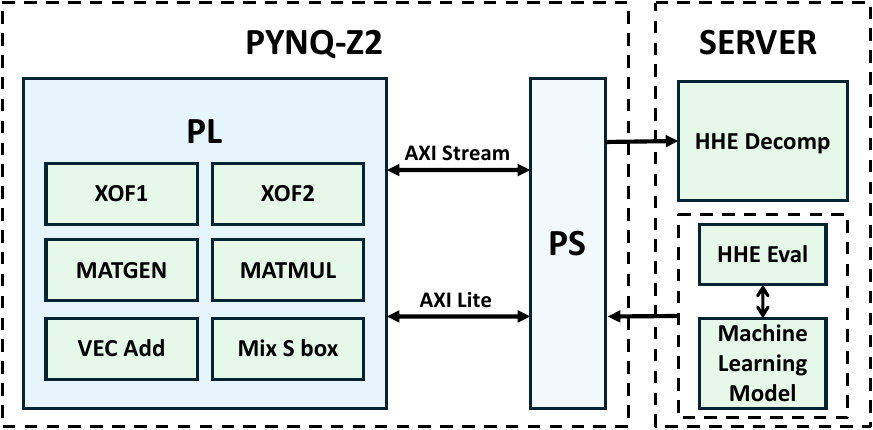}
    \caption{HHEML System Overview}
    \label{fig:System overview}
\end{figure}

\subsection{System Overview} 
The proposed HHEML framework, shown in Fig. \ref{fig:System overview}, is composed of two tightly integrated parts: a client implemented on a PYNQ-Z2 FPGA and a server executing the GuardML-based homomorphic evaluation stack. On the client, the Programmable Logic (PL) contains a hardware accelerator derived from the Pasta on Edge design and extended to support full hybrid homomorphic encryption. Key generation, symmetric encryption, and decryption are performed entirely in hardware within a unified pipeline, removing dependence on software for latency-sensitive cryptographic operations. The Processing System (PS) handles high-level orchestration, including configuration and data scheduling, and communicates with the PL via an AXI4-Stream interface with DMA support for bulk data transfers, while control and status information are exchanged through AXI-Lite.

The processing workflow starts with plaintext inputs, such as feature vectors from edge ML applications, collected by the PS and transferred to the PL accelerator. Using the Pasta cipher, the PL rapidly encrypts the data, generating ciphertexts that are directly compatible with subsequent transformation into fully homomorphic encryption (FHE) form. These ciphertexts are sent to the server through an Ethernet link, where GuardML converts them into FHE ciphertexts and performs encrypted inference on the target model. The evaluated results are then re-encoded as symmetric ciphertexts, transmitted back to the client, and decrypted in hardware to produce the final predictions.

A key architectural feature of HHEML is the use of a pipelined datapath in the PL accelerator. By reorganizing keystream generation and data masking operations, the hardware can process multiple input blocks in parallel, minimizing latency for large datasets such as batched image inputs. This improvement enhances client-side performance while maintaining seamless integration with the GuardML software stack.

\begin{figure*}
    \centering
    \includegraphics[width=1\linewidth]{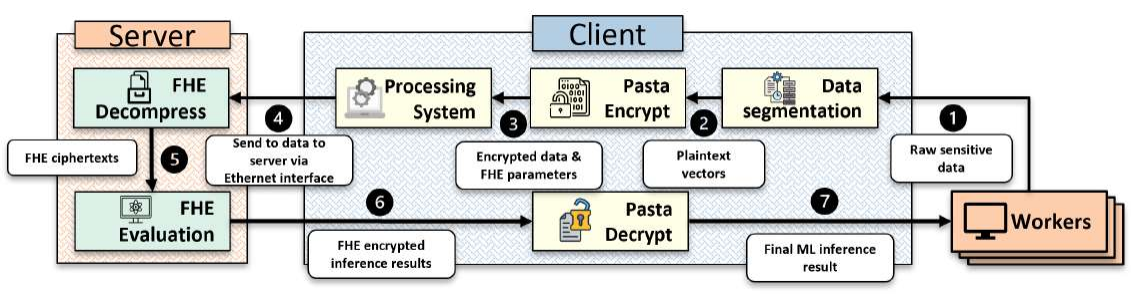}
    \caption{HHEML Workflow}
    \label{fig:placeholder}
\end{figure*}

Fig. \ref{fig:placeholder} provides an overview of this end-to-end workflow, illustrating data movement from initial plaintext capture on the client, through hardware-based encryption and network transmission, to server-side FHE evaluation, and finally secure decryption on the FPGA client.

\subsection{Modules in HHEML}
The HHEML framework operates primarily on the client side, instantiated as a tightly coupled Processing System (PS) and Programmable Logic (PL) design on the PYNQ-Z2 FPGA platform. The PL hosts a hardware implementation of the Pasta cipher based on the Pasta-4 specification, extended to support both encryption and decryption within a single, reusable AXI-Stream interface. Internally, the PL contains separate encryption and decryption modules that share common data pathways and utilize FIFOs for input and output staging, allowing seamless streaming of large data batches. A custom-designed AXI-Stream wrapper orchestrates data movement, dividing incoming plaintext or ciphertext streams into 17-word packets (32 bits each) per encryption round before injecting them into the cipher core. The same wrapper handles output staging, ensuring results are packetized and returned to the PS efficiently. Key generation and all cryptographic computations are fully implemented within the PL, allowing end-to-end symmetric processing to be completed in hardware without software intervention.

The PS in HHEML serves as the control and coordination layer. Its role is limited to configuring DMA channels, issuing start and stop commands, and supervising data flow between the host software and the hardware accelerator. It does not perform any cryptographic or preprocessing tasks, relying entirely on the PL for key scheduling, keystream generation, and encryption or decryption. Communication between PS and PL is handled via an AXI4-Stream interface combined with DMA for high-throughput data transfer, while control and status signals are managed through a lightweight AXI-Lite interface. Once data encryption is complete, the resulting ciphertexts, together with their symmetric encryption keys (protected under the SE scheme), are transmitted to the server through a single Ethernet link, where they are transformed into FHE ciphertexts and processed within the GuardML inference pipeline.

This modular hardware-software partitioning enables a clear separation of responsibilities: the PL delivers high-performance, hardware-accelerated encryption and decryption with complete key management, while the PS maintains minimal control overhead. Such a design reduces latency on the client side and provides a scalable foundation for hybrid homomorphic encryption in privacy-preserving machine learning applications.

\section{Detail Hardware Design and Optimizations} 
The hardware implementation of HHEML consists of several core components derived from the Pasta on Edge design, which serves as our baseline cryptoprocessor architecture for hybrid homomorphic encryption. Leveraging this foundation, we introduce an optimized pipelined permutation pipeline tailored for privacy‑preserving machine learning workloads.

\subsection{Baseline Hardware Flow}
We adopt the Pasta‑4 specification from the original Pasta on Edge framework, which implements key generation, keystream expansion, and data masking fully in hardware. In their design, a single SHAKE‑based XOF module generates the pseudorandom stream used to mask plaintext inputs in 17-word blocks per encryption round (each word 32 bits). An AXI‑Stream wrapper orchestrates the flow of data into and out of the Pasta core, while internal FIFOs stage input and output data to balance throughput. This architecture delivers a symmetric encryption accelerator that achieves significant speedups—on the order of 40× to over 100×—compared to CPU baselines on both FPGA and ASIC platforms 

\subsection{Processing System Design}
The Processing System (PS) in HHEML fulfills two essential roles: hardware control and network communication. On the control side, the PS configures DMA engines, issues start and stop signals to the Programmable Logic (PL), and monitors the status of encryption or decryption operations. This lightweight management approach minimizes PS computational overhead while allowing the PL to handle all cryptographic tasks autonomously.

In addition to local control, the PS maintains the communication interface between the FPGA-based client and the remote server. An Ethernet controller integrated into the PS stack manages data transfers over the network, transmitting encrypted feature vectors and their associated symmetric encryption keys (protected under the SE scheme) to the server for transformation into fully homomorphic ciphertexts. The same interface handles receiving the processed results after FHE evaluation, which are then passed to the PL for final decryption. 

\subsection{Programmable Logic Model}
Within the PL, the hardware implements the Pasta‑4 cipher core following the baseline design, including the SHAKE128‑based XOF and the permutation-based masking engine. Input and output FIFOs buffer plain/cipher text streams to accommodate batch transfers via AXI‑Stream, preserving high throughput even when transient stalls occur. Both encryption and decryption modules share the same AXI‑Stream wrapper and FIFO interface, simplifying control logic and reuse of buffers. A centralized round counter ensures correct sequencing of XOF rounds and block alignment throughout the pipeline.

\begin{table*}[htbp]
\centering
\caption{Implementation Result}
\label{t1}
\begin{tabular}{l|c|c|c|c|c|c|c|c}
\hline
Work & Clock Freq (MHz) & LUTs & FFs & DSP & BRAMs & 1-round Enc.(μs) & MNIST Enc.(μs) & Power (W)\\
\hline
Pasta on Edge\cite{aikata2025pasta} & 75& 23,736& 11,132& 64 &0 &66.1 &3,106.7& 1.2\\
\textbf{Our Work} & 75& 34,419& 18,372& 64& 2 &\textbf{34.3}&\textbf{1,553.4}&1.505 \\
\hline
\end{tabular}
\end{table*}
\begin{table*}[ht]
\centering
\caption{Throughput Improvement from Two‑XOF Pipeline}
\label{t2}
\begin{tabular}{l|c|c|c}
\hline
Configuration & Number of XOF Modules & Rounds per MNIST Image & Relative Throughput \\
\hline
Single‑XOF\cite{aikata2025pasta} & 1 & 47& 1.00$\times$ \\
\textbf{Our Work}& 2 & \textbf{24}& \textbf{1.95}$\times$\\
\hline
\end{tabular}
\end{table*}

\subsection{Data Transfer}

Data movement between PS and PL is handled via AXI4‑Stream interfaces linked to DMA controllers, enabling high-throughput packet transfers. Incoming plaintext data is streamed into PL in contiguous blocks, buffered in the input FIFO and then consumed by the Pasta core. Processed ciphertext is similarly queued in the output FIFO and transferred back to PS. This DMA-based, FIFO-buffered design minimizes PS-side data staging and avoids processing stalls, even during large ML workloads such as batch‑encryption of image datasets.

\subsection{Pipeline Optimization}

\begin{figure}[b]
    \centering
    \includegraphics[width=1\linewidth]{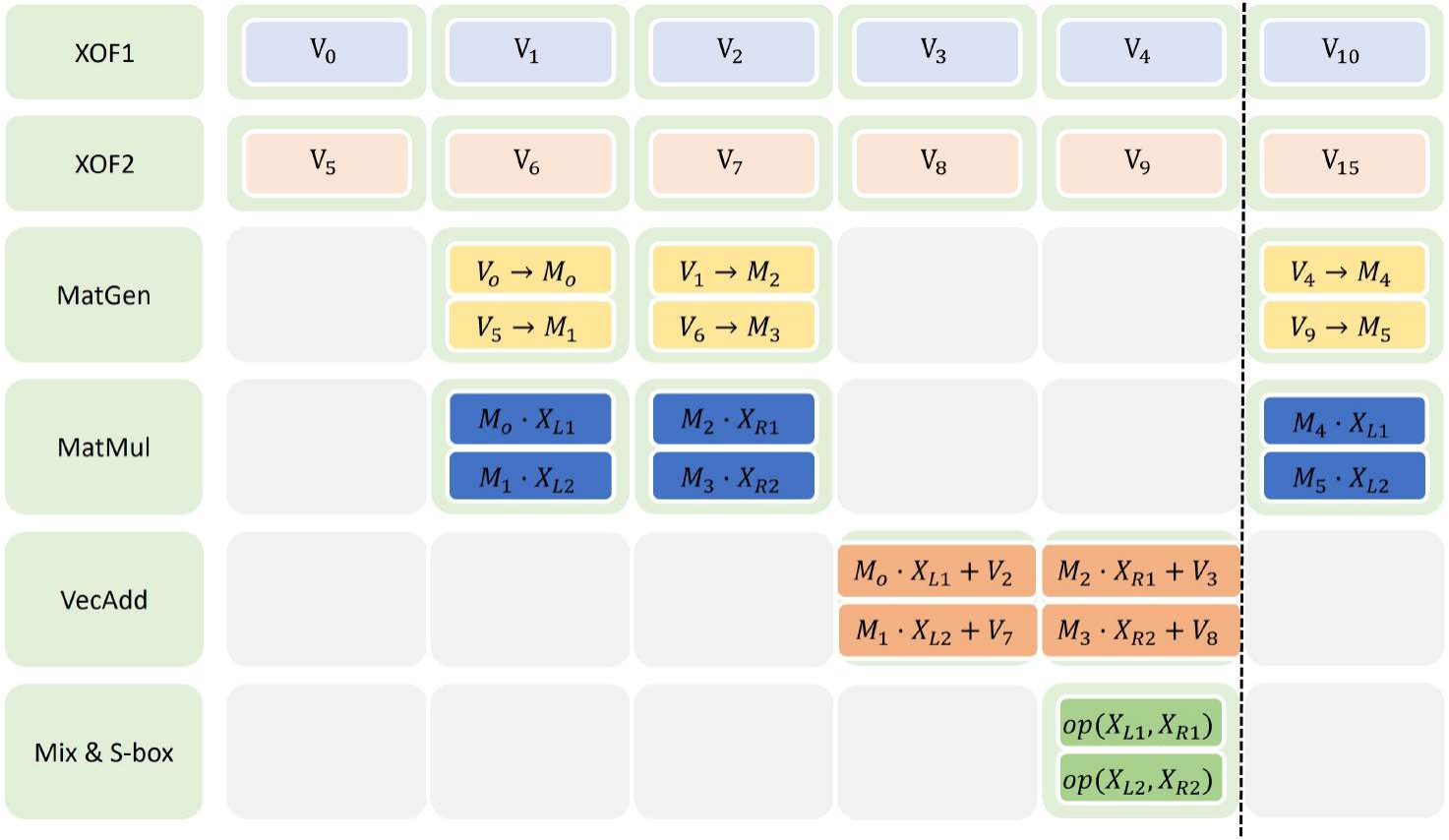}
    \caption{XOF pipeline design}
    \label{fig:XOF pipeline}
\end{figure}
Building on the baseline, HHEML introduces a two‑XOF module pipeline to greatly improve throughput for large data encryption. The entire design is shown in Fig. \ref{fig:XOF pipeline}. In contrast to the single‑module design in Pasta on Edge, our architecture replicates the SHAKE128 XOF block, allowing parallel mask generation over interleaved data streams. A centralized scheduler (round counter) dynamically assigns input words to one of two XOF modules in alternating fashion, driving both units concurrently. This architectural enhancement effectively halves the number of encryption rounds required for large inputs (e.g., MNIST images), reducing latency even though the fundamental core logic remains unchanged.

Such pipelining allows simultaneous execution of independent keystream generation tasks, leveraging parallelism to boost throughput without altering the core permutation logic. Importantly, because both XOF modules share the same seed and round schedule, synchronization remains straightforward via the centralized control logic. While the detailed synthesis results and resource trade‑offs will be discussed in Section \ref{sec:evaluation}, the design preserves correctness by maintaining ordered output and eliminating data interleaving artifacts.

By combining the Pasta on Edge baseline with this pipelined XOF architecture, HHEML achieves significantly higher encryption throughput with minimal overhead in PL resource use, making it well-suited for PPML tasks on edge devices.

\section{Evaluation}
\label{sec:evaluation}
\subsection{Evaluation Setup}
We evaluate HHEML on a Xilinx PYNQ-Z2 FPGA platform equipped with a dual-core ARM Cortex-A9 PS and 650 MHz programmable logic fabric. The client-side accelerator is clocked at 100 MHz, while the server-side computation runs the GuardML framework on an Intel i7-10750H CPU (2.6 GHz) with 16 GB RAM. All hardware designs are synthesized using Vivado 2022.1 with default timing constraints, and DMA drivers are implemented in PYNQ Python APIs for PS–PL data transfer. For all FPGA-based experiments, we use MNIST as the primary dataset (28×28 images), encoded into 784-word plaintext blocks (32-bit words).

The baseline comparison includes:

\begin{enumerate}
\item Pasta on Edge – the original FPGA implementation of the Pasta cipher, representing the prior state-of-the-art HHE hardware module without pipelining optimization.
\item GuardML (software) – a pure software implementation of HHE for PPML inference, representative of typical hybrid homomorphic approaches deployed on CPUs.
\end{enumerate}

\subsection{Results}

\begin{table*}[ht]
\centering
\caption{End‑to‑end PPML latency and communication for HHE Encrpytion}
\label{t3}
\begin{tabular}{l|c|c|c}
\hline
System & Client internal Comm (ms)&Client→Server Comm (ms)& Total Latency (ms)\\
\hline
GuardML\cite{frimpong2024guardml} &-&356& 356\\
\textbf{HHEML} & \textbf{6.5}&\textbf{0.4}& \textbf{6.9}\\
\hline
\end{tabular}
\end{table*}

\begin{table*}[ht]
\centering
\caption{End‑to‑end runtime and communication costs on different datasets}
\label{t4}
\begin{tabular}{l|c|c|c|c}
\hline
Dataset & System & Client-side (ms)&  Server-side Decomp (s)&Server-side Eval (s) \\
\hline
MNIST & GuardML\cite{frimpong2024guardml} & 356&  9,397.5&54.9\\
MNIST & \textbf{Our Work} & \textbf{6.5}&  9,397.5&54.9 \\
\hline
\end{tabular}
\end{table*}

Table \ref{t1} summarizes the hardware resource utilization and runtime comparison between HHEML and Pasta on Edge. Our design achieves nearly a 2$\times$ speedup for end-to-end MNIST encryption while maintaining comparable resource usage on the PYNQ-Z2 device. This improvement stems from introducing a second XOF module that enables parallel keystream generation, effectively reducing the number of permutation rounds required per input batch. Although the addition of an extra XOF module slightly increases logic utilization, the overall power consumption remains comparable to the original design, which was primarily evaluated in an ASIC setting. We argue that this represents a favorable trade-off between performance and resource cost. Furthermore, compared to modern CPU-based PPML solutions deployed on high-power desktop systems (typically consuming 300W or more), our FPGA-based implementation demonstrates significantly lower power requirements while delivering competitive performance.

Table \ref{t2} further quantifies the benefit of the two-XOF pipeline: large plaintext blocks (e.g., MNIST samples) that require 47 rounds under the single-XOF design are reduced to 24 rounds, effectively doubling throughput without algorithmic changes to the Pasta cipher. This optimization is particularly beneficial for machine learning workloads that process high-dimensional input vectors.

Table \ref{t3} evaluates end-to-end latency, including PS–PL DMA transfers and Ethernet communication overhead. Compared to GuardML’s software pipeline, HHEML reduces client-side runtime from hundreds of milliseconds to single-digit milliseconds, yielding a $>$50× latency improvement for LAN deployments. The server-side runtime remains identical since our system reuses GuardML’s FHE evaluation stack, but overall system latency drops significantly due to hardware acceleration of symmetric encryption.

Table \ref{t4} reports the breakdown of end-to-end runtime for the MNIST dataset when using GuardML’s software implementation versus the proposed HHEML hardware-accelerated design. The client-side processing time is significantly reduced from 356 ms to 6.5 ms due to offloading symmetric encryption to the FPGA, eliminating the high computational cost of software-based Pasta encryption. The server-side decomposition and homomorphic evaluation phases remain unchanged, as HHEML reuses the same GuardML FHE processing stack. These results highlight that the main performance bottleneck in hybrid homomorphic encryption pipelines resides on the client side, and that hardware acceleration can substantially improve overall responsiveness of privacy-preserving machine learning without requiring modifications to the server-side inference framework.

Overall, HHEML demonstrates that FPGA-based pipelining of Pasta significantly improves the practicality of HHE for PPML. Compared to both software-only frameworks and prior FPGA implementations, our design achieves lower latency, and higher throughput while preserving end-to-end cryptographic security.

\section{Conclusion}
\label{sec:conclusion}
We presented HHEML, a hardware-accelerated hybrid homomorphic encryption framework for privacy-preserving machine learning on edge devices. Built on the Pasta cipher and inspired by \textit{Pasta on Edge}, our FPGA design unifies key generation, encryption, and decryption in a PS–PL pipeline with Ethernet-based offloading to a remote FHE server.
A dual-XOF pipelined permutation architecture reduces encryption rounds for high-dimensional data, nearly doubling throughput with minimal area overhead. On a PYNQ-Z2 platform, HHEML achieves over 50$\times$ lower client-side latency and close to 2$\times$ higher throughput compared to prior FPGA-based HHE solutions.
These results highlight that lightweight FPGA acceleration makes hybrid HE practical for real-world PPML, enabling secure, low-latency inference on constrained devices. Future work will focus on multi-core pipelining and tighter integration with advanced FHE backends for further scalability and efficiency improvements.

\bibliographystyle{IEEEtran}
\bibliography{citations}

\begin{thebibliography}{10}
\providecommand{\url}[1]{#1}
\csname url@samestyle\endcsname
\providecommand{\newblock}{\relax}
\providecommand{\bibinfo}[2]{#2}
\providecommand{\BIBentrySTDinterwordspacing}{\spaceskip=0pt\relax}
\providecommand{\BIBentryALTinterwordstretchfactor}{4}
\providecommand{\BIBentryALTinterwordspacing}{\spaceskip=\fontdimen2\font plus
\BIBentryALTinterwordstretchfactor\fontdimen3\font minus \fontdimen4\font\relax}
\providecommand{\BIBforeignlanguage}[2]{{%
\expandafter\ifx\csname l@#1\endcsname\relax
\typeout{** WARNING: IEEEtran.bst: No hyphenation pattern has been}%
\typeout{** loaded for the language `#1'. Using the pattern for}%
\typeout{** the default language instead.}%
\else
\language=\csname l@#1\endcsname
\fi
#2}}
\providecommand{\BIBdecl}{\relax}
\BIBdecl

\bibitem{cryptonets2016}
N.~Dowlin, R.~Gilad-Bachrach, K.~Laine, K.~Lauter, M.~Naehrig, and J.~Wernsing, ``Cryptonets: Applying neural networks to encrypted data with high throughput and accuracy,'' in \emph{International conference on machine learning}.\hskip 1em plus 0.5em minus 0.4em\relax PMLR, 2016, pp. 201--210.

\bibitem{gilad2017lola}
R.~Gilad-Bachrach, K.~Huang, K.~Laine, K.~Lauter, M.~Naehrig, and J.~Wernsing, ``Lola: A low latency secure neural network inference framework,'' in \emph{Proceedings of the 34th International Conference on Machine Learning}.\hskip 1em plus 0.5em minus 0.4em\relax PMLR, 2017, pp. 327--336.

\bibitem{rathee2020falcon}
D.~Rathee, D.~Gupta, A.~Rastogi, and R.~Sharma, ``Falcon: Fast spectral inference on encrypted data,'' in \emph{International Conference on Advances in Cryptology}.\hskip 1em plus 0.5em minus 0.4em\relax Springer, 2020, pp. 438--467.

\bibitem{wang2021cheetah}
X.~Wang, S.~Wagh, D.~Gupta, S.~Choi, P.~Mittal, N.~Chandran, and A.~Sahai, ``Cheetah: Lean and fast secure two-party deep neural network inference,'' in \emph{USENIX Security Symposium}, 2021, pp. 1--18.

\bibitem{podschwadt2022survey}
R.~Podschwadt, D.~Takabi, P.~Hu, M.~H. Rafiei, and Z.~Cai, ``A survey of deep learning architectures for privacy-preserving machine learning with fully homomorphic encryption,'' \emph{IEEE Access}, vol.~10, pp. 117\,477--117\,500, 2022.

\bibitem{yuan2025approximate}
J.~Yuan, W.~Liu, J.~Shi, and Q.~Li, ``Approximate homomorphic encryption based privacy-preserving machine learning: a survey,'' \emph{Artificial Intelligence Review}, vol.~58, no.~3, p.~82, 2025.

\bibitem{lee2022privacy}
J.-W. Lee, H.~Kang, Y.~Lee, W.~Choi, J.~Eom, M.~Deryabin, E.~Lee, J.~Lee, D.~Yoo, Y.-S. Kim \emph{et~al.}, ``Privacy-preserving machine learning with fully homomorphic encryption for deep neural network,'' \emph{iEEE Access}, vol.~10, pp. 30\,039--30\,054, 2022.

\bibitem{frimpong2024guardml}
E.~Frimpong, K.~Nguyen, M.~Budzys, T.~Khan, and A.~Michalas, ``Guardml: Efficient privacy-preserving machine learning services through hybrid homomorphic encryption,'' in \emph{Proceedings of the 39th ACM/SIGAPP Symposium on Applied Computing}, 2024, pp. 953--962.

\bibitem{gong2024practical}
Y.~Gong, X.~Chang, J.~Mi{\v{s}}i{\'c}, V.~B. Mi{\v{s}}i{\'c}, J.~Wang, and H.~Zhu, ``Practical solutions in fully homomorphic encryption: a survey analyzing existing acceleration methods,'' \emph{Cybersecurity}, vol.~7, no.~1, p.~5, 2024.

\bibitem{samardzic2022craterlake}
N.~Samardzic, A.~Feldmann, A.~Krastev, N.~Manohar, N.~Genise, S.~Devadas, K.~Eldefrawy, C.~Peikert, and D.~Sanchez, ``Craterlake: a hardware accelerator for efficient unbounded computation on encrypted data,'' in \emph{Proceedings of the 49th Annual International Symposium on Computer Architecture}, 2022, pp. 173--187.

\bibitem{kim2023sharp}
J.~Kim, S.~Kim, J.~Choi, J.~Park, D.~Kim, and J.~H. Ahn, ``Sharp: A short-word hierarchical accelerator for robust and practical fully homomorphic encryption,'' in \emph{Proceedings of the 50th Annual International Symposium on Computer Architecture}, 2023, pp. 1--15.

\bibitem{fan2023tensorfhe}
S.~Fan, Z.~Wang, W.~Xu, R.~Hou, D.~Meng, and M.~Zhang, ``Tensorfhe: Achieving practical computation on encrypted data using gpgpu,'' in \emph{2023 IEEE International Symposium on High-Performance Computer Architecture (HPCA)}.\hskip 1em plus 0.5em minus 0.4em\relax IEEE, 2023, pp. 922--934.

\bibitem{krieger2024aloha}
F.~Krieger, F.~Hirner, A.~C. Mert, and S.~S. Roy, ``Aloha-he: A low-area hardware accelerator for client-side operations in homomorphic encryption,'' in \emph{2024 Design, Automation \& Test in Europe Conference \& Exhibition (DATE)}.\hskip 1em plus 0.5em minus 0.4em\relax IEEE, 2024, pp. 1--6.

\bibitem{dobraunig2023pasta}
C.~Dobraunig, L.~Grassi, L.~Helminger, C.~Rechberger, M.~Schofnegger, and R.~Walch, ``Pasta: A case for hybrid homomorphic encryption,'' \emph{IACR Transactions on Cryptographic Hardware and Embedded Systems}, vol. 2023, no.~3, pp. 30--73, 2023.

\bibitem{masta2024}
L.~Grassi, Q.~Guo, P.~Pessl, C.~Rechberger, and M.~Schofnegger, ``Masta: A new family of symmetric ciphers tailored for hybrid homomorphic encryption,'' \emph{IACR Transactions on Cryptographic Hardware and Embedded Systems}, vol. 2024, no.~2, pp. 1--27, 2024.

\bibitem{elisabeth2024}
C.~Bouillaguet, T.~Fuhr, M.~Naya-Plasencia, and R.~Zimmer, ``Elisabeth: Efficient lightweight symmetric cipher for bootstrappable homomorphic encryption,'' \emph{IACR Transactions on Cryptographic Hardware and Embedded Systems}, vol. 2024, no.~1, pp. 90--115, 2024.

\bibitem{sok2025}
C.~Dobraunig, L.~Grassi, L.~Helminger, C.~Rechberger, and M.~Schofnegger, ``Sok: Fhe-friendly symmetric ciphers and transciphering for secure machine learning,'' in \emph{2025 IEEE Symposium on Security and Privacy (S\&P)}.\hskip 1em plus 0.5em minus 0.4em\relax IEEE, 2025, pp. 1--20.

\bibitem{nguyen2024pervasive}
K.~Nguyen, M.~Budzys, E.~Frimpong, T.~Khan, and A.~Michalas, ``A pervasive, efficient and private future: Realizing privacy-preserving machine learning through hybrid homomorphic encryption,'' in \emph{2024 IEEE Conference on Dependable, Autonomic and Secure Computing (DASC)}.\hskip 1em plus 0.5em minus 0.4em\relax IEEE, 2024, pp. 47--56.

\bibitem{aikata2025pasta}
A.~Aikata, D.~S. Sobrino, and S.~S. Roy, ``Pasta on edge: Cryptoprocessor for hybrid homomorphic encryption,'' in \emph{2025 Design, Automation \& Test in Europe Conference (DATE)}.\hskip 1em plus 0.5em minus 0.4em\relax IEEE, 2025, pp. 1--7.

\bibitem{jeon2025presto}
Y.~Jeon, M.~Erez, and M.~Orshansky, ``Presto: Hardware acceleration of ciphers for hybrid homomorphic encryption,'' \emph{arXiv preprint arXiv:2507.00367}, 2025.

\bibitem{aikata2023reed}
A.~Aikata, A.~C. Mert, S.~Kwon, M.~Deryabin, and S.~S. Roy, ``Reed: Chiplet-based accelerator for fully homomorphic encryption,'' \emph{arXiv preprint arXiv:2308.02885}, 2023.

\bibitem{kim2024cheddar}
J.~Kim, W.~Choi, and J.~H. Ahn, ``Cheddar: A swift fully homomorphic encryption library for cuda gpus,'' \emph{arXiv preprint arXiv:2407.13055}, 2024.

\bibitem{yang2024phantom}
H.~Yang, S.~Shen, W.~Dai, L.~Zhou, Z.~Liu, and Y.~Zhao, ``Phantom: A cuda-accelerated word-wise homomorphic encryption library,'' \emph{IEEE Transactions on Dependable and Secure Computing}, vol.~21, no.~5, pp. 4895--4906, 2024.

\bibitem{tong2024accurate}
J.~Tong, J.~Dang, A.~Golder, A.~Raychowdhury, C.~Hao, and T.~Krishna, ``Accurate low-degree polynomial approximation of non-polynomial operators for fast private inference in homomorphic encryption,'' \emph{Proceedings of Machine Learning and Systems}, vol.~6, pp. 210--223, 2024.

\bibitem{podschwadt2024memory}
R.~Podschwadt, P.~Ghazvinian, M.~GhasemiGol, and D.~Takabi, ``Memory efficient privacy-preserving machine learning based on homomorphic encryption,'' in \emph{International Conference on Applied Cryptography and Network Security}.\hskip 1em plus 0.5em minus 0.4em\relax Springer, 2024, pp. 313--339.

\end{thebibliography}

\end{document}